\documentclass{emulateapj}
\usepackage{amsmath,amstext,amsgen,amsbsy,amsopn,amsfonts,theorem}
\usepackage{natbib}

\usepackage{graphicx}
\usepackage{xspace}
\usepackage{amsmath}
\usepackage{hyperref}
\usepackage{framed}
\usepackage{txfonts}
\usepackage{epstopdf}

\makeatletter
\def\mr@ignsp#1 {\ifx\:#1\@empty\else #1\expandafter\mr@ignsp\fi}%
\newcommand{\multiref}[1]{\begingroup%\let\protect\string%
\xdef\mr@no@sparg{\expandafter\mr@ignsp#1 \: }%
\def\mr@comma{}%
\@for\mr@refs:=\mr@no@sparg\do{\mr@comma\def\mr@comma{,}\ref{\mr@refs}}%
\endgroup}
\makeatother

\newcommand{\hypref}[2]{\ifx\href\asklfhas #2\else\href{#1}{#2}\fi}
\newcommand{\Secref}[1]{\S~\multiref{#1}}

\newcommand{\Figref}[1]{Figure~\multiref{#1}}
\newcommand{\figref}[1]{Fig.~\multiref{#1}}
\renewcommand{\eqref}[1]{(\multiref{#1})}

\def\[{\begin{equation}}
\def\]{\end{equation}}
\def\<{\begin{eqnarray}}
\def\>{\end{eqnarray}}

\newcommand{\nln}{\nonumber\\}

\newcommand{\NHI}{N_{\rm HI}}
\newcommand{\pcms}{{\rm ~cm}^{-2}}

\begin{document}

\title{On the Origin of the high column density turnover in the HI column density distribution}

\author{Denis Erkal\altaffilmark{1}, Nickolay Y. Gnedin\altaffilmark{2,3,4} and Andrey V. Kravtsov\altaffilmark{3,4,5}}
\altaffiltext{1}{Department of Physics, The University of Chicago, Chicago, IL 60637,
USA} \altaffiltext{2}{Particle Astrophysics Center, Fermi National Accelerator
Laboratory, Batavia, IL 60510, USA} \altaffiltext{3}{Kavli Institute for Cosmological
Physics and Enrico Fermi Institute, The University of Chicago, Chicago, IL 60637, USA}
\altaffiltext{4}{Department of Astronomy \& Astrophysics, The University of Chicago,
Chicago, IL 60637 USA} \altaffiltext{5}{Enrico Fermi Institute, The University of
Chicago, Chicago, IL 60637, USA}
\begin{abstract}
We study the high column density regime of the HI column density distribution function
and argue that there are two distinct features: a turnover at $\NHI \approx 10^{21}
\pcms$ which is present at both $z=0$ and $z\approx 3$, and a lack of systems above
$\NHI \approx 10^{22} \pcms$ at $z=0$. Using observations of the column density
distribution, we argue that the HI-H2 transition does not cause the turnover at $\NHI
\approx 10^{21} \pcms$, but can plausibly explain the turnover at $\NHI \gtrsim
10^{22} \pcms$. We compute the HI column density distribution of individual galaxies
in the THINGS sample and show that the turnover column density depends only weakly on
metallicity. Furthermore, we show that the column density distribution of galaxies,
corrected for inclination, is insensitive to the resolution of the HI map or to
averaging in radial shells. Our results indicate that the similarity of HI column
density distributions at $z=3$ and $z=0$ is due to the similarity of the maximum HI
surface densities of high-z and low-z disks, set presumably by universal processes
that shape properties of the gaseous disks of galaxies. Using fully cosmological
simulations, we explore other candidate physical mechanisms that could produce a
turnover in the column density distribution. We show that while turbulence within GMCs
cannot affect the DLA column density distribution, stellar feedback can affect it
significantly if the feedback is sufficiently effective in removing gas from the
central 2-3 kpc of high-redshift galaxies. Finally, we argue that it is meaningful to
compare column densities averaged over $\sim$ kpc scales with those estimated from
quasar spectra which probe sub-pc scales due to the steep power spectrum of HI column
density fluctuations observed in nearby galaxies.
\end{abstract}

\section{Introduction}

The statistical distribution of HI column densities is one of the most widely used
statistics to describe the statistical properties of atomic hydrogen absorption
systems seen in quasar spectra over a wide range of redshifts. Recently,
high-resolution HI maps of nearby galaxies have been used to study HI column density
distributions of both individual galaxies and statistically for the entire galaxy
population \citep[e.g.,][]{2006ApJ...643..675Z}. Current measurements of the column
density distribution span almost ten orders of magnitude in column density, $\NHI$,
ranging from the highly ionized gas in the intergalactic medium (IGM), $\NHI \sim
10^{12} - 1.6 \times 10^{17}\pcms$, to the predominantly neutral gas associated with
the interstellar medium (ISM) of galaxies and their satellites, $\NHI > 10^{19}\pcms$
\citep[e.g.,][]{2010ApJ...718..392P}. Since physically diverse systems contribute to
the distribution, a theoretical understanding of the entire HI column density
distribution requires understanding IGM gas as well as gas within galaxies and the
circumgalactic medium. As such, the HI column density distribution is a key test of
the CDM structure formation paradigm.

Initial measurements of the HI column density distribution were consistent with a
single power-law function over the entire column density range
\citep[e.g.,][]{1987ApJ...321...49T}. However, subsequent surveys have found
considerably more structure: e.g., flattening at $\NHI \approx 10^{16}-10^{18} \pcms$
 \citep{1993MNRAS.262..499P}, and steepening of the function at higher
column densities corresponding to the damped Lyman $\alpha$ (DLA; $\NHI > 2\times
10^{20}\pcms$) systems, and in particular, pronounced steepening  at $\NHI \gtrsim
10^{21} \pcms$
\citep{2000ApJ...543..552S,2005ApJ...635..123P,2005ARA&A..43..861W,2009A&A...505.1087N}.
Most recently, \citet{2010ApJ...718..392P} found that the HI column density
distribution has six different regimes, each approximated by its own power law, which
intersect at column densities $\NHI = [10^{14.5},10^{17.3},10^{19},
10^{20.3},10^{21.75}] \pcms$.

This rich structure is presumably the result of physically distinct populations of
absorbers. At the lowest column densities of the Lyman $\alpha$ forest, $\NHI\approx
10^{12} - 10^{17} \pcms$, absorbers arise in the highly ionized intergalactic gas and
both the shape of the column density distribution and its evolution with redshift are
now fairly well understood \citep[e.g., see][for a recent review]{meiksin09}. At
column densities of $\NHI \approx 10^{17}-10^{20} \pcms$ the absorbers arise in Lyman
Limit Systems (LLS). These systems are thought to arise in circumgalactic gas and
clouds \citep[e.g.,][]{fumagalli_etal11}  and are capable of self-shielding against
ionizing radiation \citep{2011ApJ...737L..37A}, which results in a pronounced
flattening of the distribution at these column densities. At the highest column
densities of $\NHI\gtrsim 10^{20} \pcms$, DLA absorption lines are thought to arise in
the interstellar medium of high-redshift galaxies
\citep[e.g.,][]{2005ARA&A..43..861W}. These absorption lines thus directly probe
properties of cold gas in high-redshift disks which contain most of the cold gas in
the universe.

As we discuss in the next section, modeling the column density distribution and other
properties of these DLA systems is both challenging and interesting. It is challenging
because the small-scale density and velocity structure of the high-redshift ISM must
be reproduced correctly. This same reason makes it interesting since the column
density distribution provides a unique way of testing the small-scale gas distribution
in theoretical models.

The column density distribution for DLAs has two important features which we will
discuss in this paper. First, at both low and high redshifts there is a pronounced
turnover at $\NHI \approx 10^{21} \pcms$
\citep{2005MNRAS.364.1467Z,2009A&A...505.1087N,2009ApJ...696.1543P}. The location of
this turnover is found by fitting the column density distribution with either a gamma
distribution or a double power-law. At low redshift, where observations extend to
higher column density, there is an additional feature. We see that there is a
transition from HI-H$_2$ at $N_{\rm{H}} \approx 10^{22} \pcms$ which leads to a lack
of HI systems above this column density
\citep{2005MNRAS.364.1467Z,2006ApJ...643..675Z}. At high redshift this region has not
been probed due to insufficient statistics. Previously, these two features were not
distinguished but in this paper we will treat them separately since we are interested
in differentiating the physical mechanisms which control each of them. In previous
works, authors tried to understand the steepening in the column density distribution
at high column densities, $\NHI > 10^{21}\pcms$. Proposed explanations discussed so
far include selection effects due to dust obscuration, conversion of atomic hydrogen
to molecular hydrogen, and inclination effects due to randomly oriented galaxies.

Dust obscuration was considered by \citet{fall_pei93} and \citet{2005A&A...444..461V},
who argued that it could explain the steepening in the DLA regime. The importance of
dust obscuration is still a subject of debate
\citep{2001A&A...379..393E,2006ApJ...646..730J,2010MNRAS.406.2235F,2010PASP..122..619K}.
In particular, \citet{2010PASP..122..619K} and, most recently,
\citet{2011arXiv1109.4225K} found that a non-negligible fraction of high-metallicity
DLAs are significantly reddened. Given that optical quasar samples are constructed
using color-based candidate selection, the samples may be biased against significantly
reddened QSOs. In the extreme case, when a QSO is completely obscured in the UV and
optical range by dust, such a quasar would be missing in radio-selected samples as
well because no optical spectrum or redshift would be measured for the radio source.
However, we note that although empty-fields have been found in radio-selected samples,
these empty-fields were found to contain extended structure in the IR indicating they
were not high-redshift quasars \citep[see][for more details]{2006ApJ...646..730J}.

A physical explanation for the lack of high column density systems above $\NHI \sim
10^{22} \pcms$ discussed by a number of authors is the atomic-to-molecular transition
\citep{2001ApJ...562L..95S,2005MNRAS.356.1529H,2006ApJ...643..675Z,2009ApJ...701L..12K,2010arXiv1010.5014C,2011ApJ...737L..37A}.
The idea is that at large column densities the gas is self-shielded and hence can form
H$_2$ if there is sufficient dust. The transition from HI to H$_2$ is expected to occur
at a characteristic column density and can thus introduce a feature such as a steepening
of the HI column density distribution.

Inclination effects were first considered in
\cite{1988A&A...202L...9M,fall_pei93,1995ApJ...454..698W} where the authors considered
the column density distribution of randomly oriented disks. They assumed that the
disks had a radial column density profile which was monotonically decreasing with some
maximum column density at some minimum radius. They considered a universe populated
with these disks at random orientations and found that the column density distribution
of such a model has a kink at the maximum column density of the radial profile. We
will further generalize this model in \Secref{sec:inclination}.

In this paper we emphasize the two features in the column density distribution for
DLAs. The lack of systems in the local universe above $\NHI \approx 10^{22} \pcms$ can
be explained by the HI-H$_2$ transition \citep{2006ApJ...643..675Z}. However, we will
argue that observations show that the turnover at $\NHI \approx 10^{21} \pcms$ cannot
be explained by the same mechanism. This is because the HI-H$_2$ transition depends on
the metallicity and UV radiation of the environment. However, observations show that
the turnover at $\NHI \approx 10^{21} \pcms$ occurs both in local galaxies, which have
metallicities and interstellar UV fields close to those of the Milky Way, and in
high-redshift galaxies, which have low metallicities and high UV fluxes. The HI-H$_2$
explanation would predict a transition at a much higher column density at high
redshift and therefore a turnover which is not independent of redshift. In addition,
we show that the HI column density distribution of individual $z\approx 0$ galaxies
all have a similar turnover which does not exhibit a significant dependence on
metallicity. Furthermore, we can show that the turnover at $\NHI \sim 10^{21} \pcms$
can naturally arise from randomly oriented galaxies.

We present details of the argument against molecular hydrogen formation being the
cause of the turnover at $\NHI \sim 10^{21}\pcms$ in \Secref{sec:observation_section}.
In \Secref{sec:inclination} we consider the effect of randomly oriented galaxies on
the column density distribution and find that they naturally give rise to a turnover
in the column density distribution which depends only on the characteristic maximum
column density of a galaxy and not on the small-scale features. In
\Secref{sec:simulation_section} we use the column density distribution function of
DLAs as a stringent test of the gas distribution in galaxy formation simulations and
show that our simulations are discrepant with observational results. In
\Secref{sec:comparison_section} we compare our results to other recent results from
galaxy formation simulations by different groups and argue that although current
models are quite successful in matching and explaining the column density distribution
of HI absorbers over a wide range of column densities, simulation results are
generally discrepant or in tension with observations beyond the turnover at
$\NHI\gtrsim 10^{21} \pcms$.

%----------------------------------------------------------------------------
\section{Turnover in the HI column density distribution and the HI-H$_2$ transition}
\label{sec:observation_section}
%-------------------------------

\subsection{HI Column Density Distribution}
\label{sec:background_cdd}

The column density distribution of quasar absorption systems is defined as the number
of atomic hydrogen systems, $\mathcal{N}$, along a random line of sight per unit
column density, $d\NHI$, per unit absorption length, $dX$:
\< f(\NHI,z) = \frac{d^2 \mathcal{N}}{d\NHI dX}, \>
where
\< dX = \frac{H_0}{H(z)}(1+z)^2 dz .\>
In this study we focus on the highest column density regime of DLAs, which are
associated with the ISM of galaxies hosted by dark matter halos of different mass,
$M$. Using the comoving number density of halos per unit halo mass, $\partial
n(M,z)/\partial M$, along with the differential HI cross-section associated with such
a halo, $\partial\sigma(M,\NHI,z)/\partial \NHI$, where $\sigma(M,\NHI,z)$ is the
cross-section in proper units for producing absorbers with column densities lower than
$\NHI$, we can re-express the column density distribution as
\< f(\NHI,z) = \frac{c}{H_0} \int   \frac{\partial \sigma(M,\NHI,z)}{\partial \NHI}
\frac{\partial n(M,z)}{\partial M} dM . \label{eq:intCDD}\>
This convention emphasizes the physical factors which affect the column density
distribution function. The halo number density depends on the physics of halo
formation which is fairly well understood. However, the differential cross-section in
a galaxy with a given halo mass is determined by the surface density distribution on
the length scale at which it is being measured. For quasar absorption studies at high
redshift, this scale is comparable to the transverse physical size of the quasar
emitting region (i.e., $\ll$~pc). This distribution will be affected by ISM processes
such as turbulence, gravitational instability, gas chemistry, star formation, stellar
feedback, etc. Thus, the column density distribution provides a unique high-resolution
window into the structure of the ISM of high-redshift galaxies and the processes that
shape it.

At the same time, this presents an obvious challenge to the models since the highest
resolution galaxy formation simulations have resolutions which are orders of magnitude
larger than the scale probed by quasars. Thus, when simulation results are compared to
observations, it is implicitly assumed that disparity in scales and unresolved
small-scale structure of the ISM does not affect the column density distribution.
Although highly non-trivial, as we discuss below this assumption is supported by
observations of the column density distribution at different scales in nearby galaxies
and is due to the steep power spectrum of HI column density maps.

\subsection{Observed Column Density Distribution of DLAs at high and low redshifts}
\label{sec:observational_tension}

\begin{figure}[t]
\includegraphics[width=8cm]{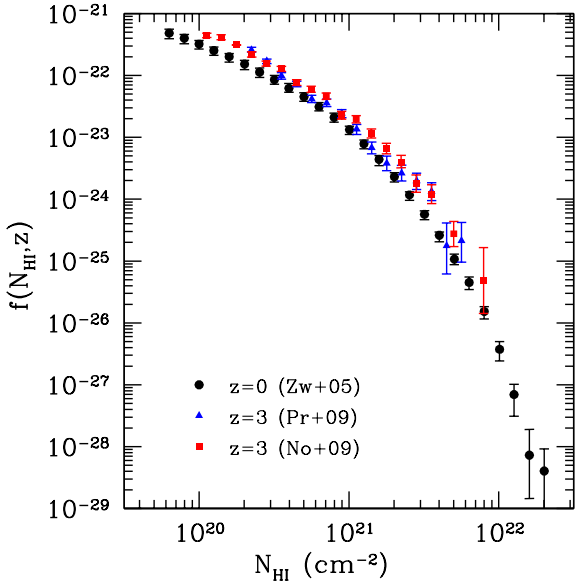}
\caption{Column density distribution for DLAs at two different redshifts. The circles,
$z=0~~(\rm{Zw}+05)$, are from observations of local galaxies based on 21cm emission
\citep{2005MNRAS.364.1467Z}. The triangles, $z=3~~(\rm{Pr}+09)$, are from observations of quasar
absorption lines in SDSS DR5 in the redshift range $z \in [2.2,5.5]$ \citep{2009ApJ...696.1543P}. Finally, the squares
$z=3~~(\rm{No}+09)$, are from observations of quasar
absorption lines in SDSS DR7 in the redshift range $z \in [2.2,5.2]$ \citep{2009A&A...505.1087N}. The decrease in the overall normalization is thought to be due to
star formation fueled by DLAs and due to processes which remove HI gas from
galaxies such as AGN activity, galactic-scale winds, tidal effects, and ram pressure
stripping \citep[see][]{2009ApJ...696.1543P}.}
\label{fig:zwaan_vs_prochaska_vs_noterdaeme}
\end{figure}

In \Figref{fig:zwaan_vs_prochaska_vs_noterdaeme}, we plot the column density
distribution of systems in the local universe and at $z=3$. As was noted before
\citep{2005MNRAS.364.1467Z,2005ApJ...635..123P,2009A&A...505.1087N}, the column
density distribution of DLAs at $z\approx 0$ and $z\sim 3$ are remarkably similar and
exhibit a turnover at $\NHI \gtrsim 10^{21}\pcms$. At low redshifts, where systems
have been measured to higher column densities, this turnover continues to higher
column densities and DLAs with $\NHI\gtrsim 10^{22}\pcms$ become exceedingly rare. Now
we will discuss the physical mechanisms that may be responsible for the steep decrease
of the cross-section of high column density gas.

Column densities above $\NHI \sim 10^{22}\pcms$ are typical for giant molecular clouds
(GMCs) and $f(\NHI,z)$ in this regime should probe gas associated with, or fueling
formation of, GMCs and star formation in galaxies. For this reason, it is natural to
associate the lack of systems above $\NHI \sim 10^{22} \pcms$ with the HI-H$_2$
transition. Indeed, \cite{2006ApJ...643..675Z} showed that for local $z\approx 0$
galaxies the column density distribution of HI smoothly joins onto the column density
distribution of molecular gas, with a cross-over at $N_{\rm{H}} \sim 10^{22} \pcms$,
indicating that the turnover is due to the HI-H$_2$ transition. Recently, the HI-H$_2$
transition was invoked in theoretical studies as a significant factor in steepening the
column density distribution at $\NHI \gtrsim 10^{22}\pcms$ even for high-$z$ DLAs
\citep{2010arXiv1010.5014C,2011ApJ...737L..37A}. However, the H$_2$ models used in these
studies were calibrated on local galaxies, most of which have solar or super-solar
metallicities. The HI-H$_2$ transition, on the other hand, is expected to depend on both
the amount of dust and the ambient far UV radiation \citep[e.g.][]{1993ApJ...411..170E}.
The dust helps to shield molecular gas from photodissociation and acts as a catalyst for
H$_2$ formation. If the dust-to-gas ratio is proportional to the metallicity of the gas,
as indicated by observations at $Z\gtrsim 0.1Z_{\odot}$ \citep[e.g.,][]{draine_etal07},
the HI-H$_2$ transition will correspondingly depend on the metallicity of the absorbing
gas. In \cite{2001ApJ...562L..95S}, the author developed a theoretical model which
predicted the maximum HI column density as a function of metallicity. However, his model
predicted that the maximum HI column density scales as $\approx 1/\sqrt{Z}$, in
disagreement with local observations of the HI-H$_2$ transition.

\cite{2006ApJ...636..891G} measured the transition in the Milky Way and
\cite{2002ApJ...566..857T} measured the transition in the Small Magellanic Cloud (SMC)
and Large Magellanic Cloud \citep[LMC; see also a compilation of archival measurements in
the SMC, LMC and Milky Way by][]{welty_etal11}. The transition in the Milky Way occurs at
a column density of $N_H \approx 10^{20.4} - 10^{20.7}\pcms$, depending on whether the
line of sight is at high latitude or along the disk. This difference is expected due to a
higher H$_2$-dissociating UV field in the disk plane. The transition in the LMC
($Z\approx 0.3Z_{\odot}$) occurs at $N_H \geq 10^{21.3}\pcms$ and the transition in the
SMC ($Z\approx 0.15Z_{\odot}$) occurs at $N_H \geq 10^{22}\pcms$. Thus, the transition
column density in these galaxies scales as $\approx 1/Z$ or even more strongly.

In light of the expected metallicity dependence of the HI-H$_2$ transition, the
similarity of the low and high-$z$ HI column density distribution turnover at $\NHI \sim
10^{21}\pcms$ in \Figref{fig:zwaan_vs_prochaska_vs_noterdaeme} is remarkable and strongly
disfavors the HI-H$_2$ transition as the physical origin of the turnover. While the
low-redshift systems have metallicities of $[Z/H] \sim -0.5$ \citep{2005MNRAS.364.1467Z},
high-redshift DLAs have typical metallicities of $[Z/H] \lesssim -1$
\citep{2009A&A...505.1087N,2009ApJ...696.1543P}, and we would thus expect the turnover in
the high-$z$ $f(\NHI,z)$ to occur at least 0.5 dex higher in column density. Yet, the
column density distributions of these systems exhibit turnovers at similar column
densities. This indicates that the origin of the turnover is not due to the phase
transition from atomic to molecular gas, but rather due to the overall density structure
of interstellar gas in galaxies. Remarkable similarity of the low- and high-redshift
distribution indicates that some universal and metallicity independent process is likely
responsible for this structure.

This result is consistent with the distribution of DLAs in the $\NHI$-$Z$ plane which has
been used to argue that there is a metallicity dependent maximum $\NHI$
\citep[e.g.][]{2009ApJ...701L..12K}. In \cite{2009ApJ...701L..12K}, the authors present a
compilation of observations of high-redshift DLAs and show that the distribution of DLAs
in the $\NHI$-$Z$ plane is consistent with being bounded by a metallicity dependent
maximum $\NHI$ which is due to the HI-H$_2$ transition. Binning this data in $\NHI$ gives
the column density distribution of the compiled sample which displays a turnover at $\NHI
\sim 10^{21} \pcms$. This binning exercise can be used to show that the feature is
primarily due to the density of points in the $\NHI$-$Z$ plane rather than the cutoff in
the plane due to the HI-H$_2$ transition, and hence that the turnover at $\NHI\sim
10^{21}\pcms$ is not primarily due to the HI-H$_2$ transition.

\subsection{Column Density Distribution within the ISM of $z\approx 0$ Galaxies in the THINGS Sample}
\label{sec:THINGS}

If the shape of the overall HI column density distribution is related to some
universal features of the HI surface density distribution in individual galaxies, it
would be interesting to explore whether such universality is indeed observed in
individual galaxies. To this end, we use galaxies from the THINGS sample
\citep{2008AJ....136.2563W}, in which HI surface density maps are measured for $34$
local galaxies using $21$-cm observations. The physical resolution of the maps ranges
from $\sim 100$ to $\sim 500$ pc depending on the distance to the galaxy. We use the
robust weight data cubes of the galaxies since they have the highest resolution. We
have omitted three galaxies from consideration: M81 DwA, NGC 3031, and IC 2574. M81
DwA does not have metallicity information\footnote{M81 DwA is a low-mass dwarf
irregular galaxy and, as such, likely has a low metallicity of $[Z/H]\sim -1$.
Nevertheless, HI column density within this galaxy does not exceed $\NHI\approx
5\times 10^{20}\pcms$ \citep{walter_etal07}.}, the NGC 3031 map suffers from
significant point source contamination, and IC 2574 was not available online. The SMC
column density distribution comes from a HI map of the SMC with a $200$ pc resolution,
kindly provided by Alberto Bolatto \citep[see][for details]{bolatto_etal11}.

Note that we do not smooth these maps to achieve a uniform spatial resolution
independent of the distance to each galaxy. We do not believe this will significantly
affect our results. \citet{2005MNRAS.364.1467Z} showed that the HI column density
distribution of nearby galaxies at $\NHI\sim 10^{20}-10^{22}\pcms$ did not change
significantly when they varied the resolution of their maps from 1.5 kpc to 3 kpc. In
addition, studies of nearby galaxies have shown that the HI power spectrum has a
power-law behavior over a wide range of scales, ranging from $\sim 0.1$ pc up to $\sim
10$ kpc
\citep[][etc.]{1999MNRAS.302..417S,2001ApJ...548..749E,2003A&A...411..109M,2009MNRAS.398..887D}.
These power-laws are steep with exponents ranging from $\sim -1.5$ to $\sim -3.5$.
Such a steep power-law behavior indicates that the HI maps do not have large-amplitude
fluctuations on small scales and hence the column density distribution should not look
very different as we change the scale. If the HI column density distribution of
high-redshift galaxies is characterized by similarly steep power spectra, this would
validate the assumption that the column density distribution computed from quasar
lines of sight can be meaningfully compared to the column density distribution
computed from 21-cm HI maps that probe column densities averaged on vastly larger
scales.

For each THINGS galaxy that we use, we compute the column density distribution:
\< f(\NHI) = \frac{\Delta \mathcal{N}(\NHI)}{\Delta \NHI\Delta X},
\label{eq:discretef}\>
where $\Delta \mathcal{N}(\NHI)$ is the number of resolution elements with column
density in the range $\NHI$ to $\NHI + \Delta \NHI$ and $\Delta X$ is the absorption
length for each galaxy, which is arbitrary. In terms of the formalism from
\Secref{sec:background_cdd}, $f(\NHI)$ is proportional to the differential
cross-section of the galaxy,
\< f(\NHI) \propto \frac{c}{H_0} \frac{\partial \sigma(M,\NHI)}{\partial \NHI} .\>
Since we are just comparing the differential cross-sections and shapes of the column
density distributions in different galaxies, we have adjusted the normalization by
choosing $\Delta X$ such that all column density distributions coincide with the
column density distribution of local galaxies from \cite{2005MNRAS.364.1467Z} at $\NHI
= 10^{20}\pcms$ .

\begin{figure}[t]
\includegraphics[width=8cm]{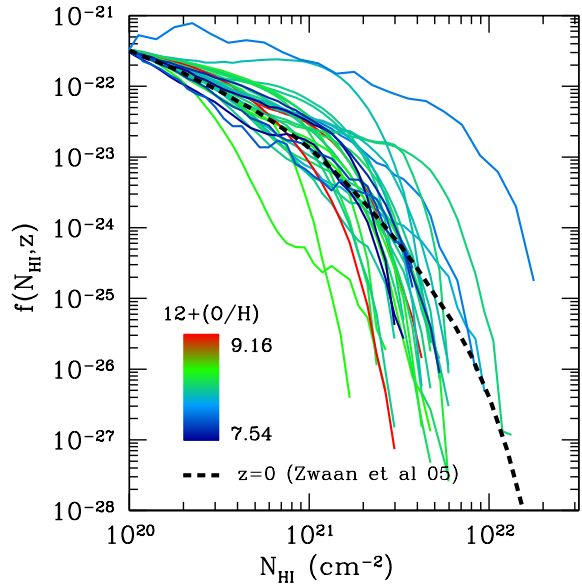}
\caption[]{Column density distribution for individual galaxies (different colored lines)
in the THINGS sample and the SMC. Each column density distribution line is colored by
the oxygen abundance of the galaxy relative to the Sun ($12+(O_\odot/H_\odot) = 8.6$),
where the color bar is linear in oxygen metallicity. For comparison, the dashed line
is the HI column density distribution of galaxies in the local universe measured by
\citet{2005MNRAS.364.1467Z}. The SMC column density distribution is the one which extends
to the highest column densities. It is constructed from a HI map of the SMC with a
$200$ pc resolution, kindly provided by Alberto Bolatto \citep[see][for details]{bolatto_etal11}.
The SMC has a metallicity of $12+(O/H)=7.96$.
The column density distribution of each galaxy has been normalized such that it
coincides with the distribution of \citet{2005MNRAS.364.1467Z} at $\NHI=10^{20}\pcms$.
Note that the column densities in this figure have not been corrected for inclination
effects because we are emphasizing the comparison with the column density distribution of
\cite{2005MNRAS.364.1467Z} which considered a statistical sample of galaxies. We will discuss
inclination effects further in \Secref{sec:inclination}.} \label{fig:THINGS_cdd_compare}
\end{figure}

The resulting column density distributions for the THINGS galaxies are plotted in
\Figref{fig:THINGS_cdd_compare}, in which each colored line represents the
distribution of an individual galaxy with color indicating the galaxy's metallicity
\citep[taken from][]{2008AJ....136.2563W}, as shown by the color bar. Although the
THINGS sample consists of a wide range of systems, from dwarf irregulars to spirals,
with metallicities varying by more than $1.5$ dex, the HI column density distributions
of these galaxies are quite similar and the scatter is surprisingly small. There is no
clear correlation of the shape of the distribution with metallicity; all galaxies have
a turnover at column density $\NHI\sim {\rm few}\times 10^{21}\pcms$.

\begin{figure}[t]
\includegraphics[width=8cm]{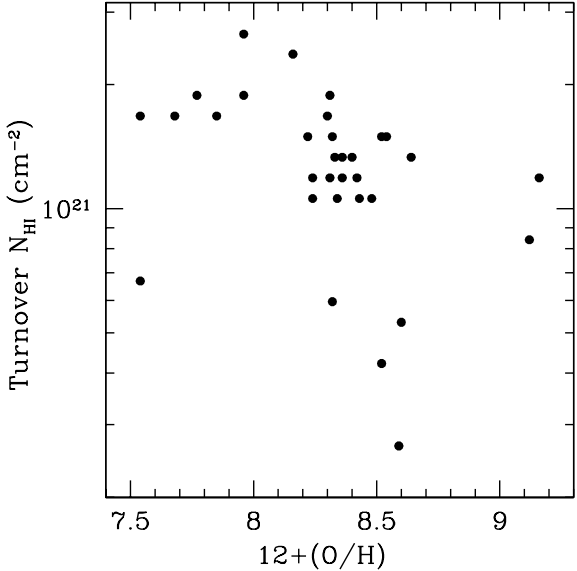}
\caption{Turnover column density in individual
THINGS galaxies  versus metallicity of the galaxy.
The turnover column density is determined by fitting a double power-law to the
deprojected (face-on) column density distribution of each galaxy. To account for
inclination, the column densities have been scaled by $\cos\theta$, where $\theta$ is
the inclination angle of the galaxy.
The correlation coefficient is $-0.40 \pm 0.14$ where the
error bars are from 10000 bootstraps over the data. There is no strong correlation of
turnover column density with metallicity, as would be expected if the HI-H$_2$ transition
was the main factor shaping the turnover of the column density distribution. Note that
discreteness of the turnover column density is 0.05 dex which is the resolution we used
to compute the column density distribution of each galaxy.}
\label{fig:THINGS_turnover_vs_met}
\end{figure}

This is shown more quantitatively in \Figref{fig:THINGS_turnover_vs_met}, where we
plot the turnover column density versus galaxy metallicity. We determine the turnover
column density by performing a double power-law fit to the deprojected (face-on)
column density distribution of each galaxy; we define the turnover column density as
the column density where the two power laws intersect.  To deproject the column
density maps we just rescale the column densities in individual pixels by
$\cos\theta$, where $\theta$ is the inclination angle of a particular galaxy. The
figure shows that there is at best only a weak correlation between turnover column
density. The correlation coefficient for the shown points is $-0.40 \pm 0.14$, where
the error is computed from 10000 bootstraps. Although a weak correlation is present,
we do not observe any strong correlation which would be expected if the turnover was
caused by the HI-H$_2$ transition. The weak correlation that may be present could be
due to the fact that the column density distribution depends weakly on the mass of the
galaxy, which would also imply a weak correlation with metallicity due to the
metallicity-stellar mass relation. In addition, we note that the two galaxies with the
highest characteristic column densities in \Figref{fig:THINGS_turnover_vs_met} (SMC,
NGC 1569) are galaxies showing signs of strong tidal interactions with their neighbors
and have associated bursting star formation. Their higher than average column
densities may thus be due to the tidally-induced inflow of gas into their central
regions.

As a check of this weak correlation, we performed two additional tests. First, we did
the analysis in \Figref{fig:THINGS_turnover_vs_met} using more recent metallicities
for the 23 galaxies in THINGS which have updated metallicities in
\cite{2010ApJS..190..233M}. We used the metallicities derived using the empirical
calibration from \citep{2005ApJ...631..231P} and found that the correlation function
became $-0.39 \pm 0.14$, indicating that the effect of using the updated metallicities
is small. Second, we performed a KS test on the inclination corrected column density
distributions. We separated the galaxies into quartiles based on their metallicities
and then performed KS tests between all pairs of galaxies with one drawn from the
bottom quartile and the other drawn from the top quartile. The result was that $90\%$
of such pairs had a KS probability greater than 0.01 of being drawn from the same
distribution. Thus, in addition to the result in \Figref{fig:THINGS_turnover_vs_met},
we have a non-parametric test which shows that the column density distribution of the
THINGS galaxies show little metallicity dependence.

The lack of strong correlation between the turnover column density and the gas
metallicity shown in \Figref{fig:THINGS_cdd_compare} and
\Figref{fig:THINGS_turnover_vs_met} is in contrast with the strong correlation that
would be expected if the HI-H$_2$ transition was shaping the turnover. In this
respect, we can also note that although the column density distribution of molecular
hydrogen extends the HI distribution smoothly in $z \approx 0$ galaxies
\citep{2006ApJ...643..675Z}, H$_2$ only affects the shape of the total hydrogen
distribution for $\NHI\gtrsim 10^{21.7}\pcms$ and $f(\NHI)\lesssim 10^{-26}$. This
limit comes from computing how many HI and H$_2$ lines of sight there are above a
given column density, which is just the integral of the column density distribution.
Above the column density of the turnover, $\NHI \sim 10^{21} \pcms$, there are more
than 10 times as many HI lines of sight as H$_2$ lines of sight.

With this argument in mind, we see that the HI-H$_2$ transition does not cause the
turnover of the column density distribution at $\NHI \sim 10^{21} \pcms$ in the local
universe. Furthermore, although the HI-H$_2$ transition can help explain the paucity
of HI absorbers with $\NHI>10^{22}\pcms$ for relatively metal-rich galaxies at
$z\approx 0$, such an explanation would not work for higher-redshift DLAs. These DLAs
have considerably smaller metallicities and for low metallicities the transition is
expected to occur at much higher column densities (see observations of the transition
in the Milky Way, SMC, and LMC discussed in \Secref{sec:observational_tension} and
theoretical predictions shown in \figref{fig:diff_ISM_comparison} below). This
reasoning predicts a tail out to high $\NHI$ in the high-redshift column density
distribution which is not present in the local universe.

The similarity of the column density distributions of the galaxies in THINGS, despite
a wide range of metallicities, sizes, star formation rates, total HI masses, and
absolute luminosities \citep{2008AJ....136.2563W}, implies that the shape of the
column density distribution at high column densities may be due to some universal
physical mechanism. In the next section we will explore the connection between the
column density distribution of a single galaxy and the column density distribution
measured for a statistical sample.

\section{Column Density Distribution of Randomly Oriented Galaxies}
\label{sec:inclination}

In the previous section, we presented arguments that the HI-H$_2$ transition cannot
explain the turnover at $\NHI \sim 10^{21} \pcms$. Here we investigate the origin of
this turnover by examining how the column density distribution within individual
galaxies contributes to the overall statistical column density distribution probed by
quasar lines of sight. The inclination angle of a galaxy can affect the column density
distribution since it affects both the number of absorption lines through a galaxy and
the column density of those lines due to projection.

The effect of inclination angle was investigated in studies by
\cite{1988A&A...202L...9M}, \citet{fall_pei93}, and \citet{1995ApJ...454..698W}, in
which  the authors considered column density distributions arising from randomly
oriented thin disks with fixed radial surface density profiles. They assumed that the
radial column density profiles are monotonically decreasing, with a maximum column
density at some minimum radius. Under these assumptions, they showed that the column
density distribution has a kink at the maximum column density of the galaxy viewed
face-on. Above this column density, all sight lines are due to projection effects and
the column density distribution behaves as a power-law with an exponent of $-3$,
independent of the specific radial profile. Below the critical column density, the
column density distribution depends on the details of the radial profile.

First, we would like to note that this result can be extended to surface density
profiles that are not axially symmetric. To prove this we consider a galaxy which,
when viewed face-on, has a column density distribution given by $f_\bot(\NHI^\bot)$,
where $\NHI^\bot$ is the column density as measured for the face-on galaxy. Next, we
rotate this galaxy so it has an inclination angle $\theta$ with respect to the
observer and consider the column density distribution, denoted by
$f_\theta(\NHI^\theta)$. We make the assumption that the galaxy is thin in the $z$
direction which allows us to relate the inclined column density, $\NHI^\theta$, to the
face-on column density, $\NHI^\bot$, via $\NHI^\theta = \NHI^\bot/ \cos\theta$.
Furthermore, when we incline the galaxy, the number of lines of sight piercing the
galaxy decreases by $\cos\theta$ due to projection effects. Therefore, the number of
lines of sight between $\NHI^\theta$ and $\NHI^\theta + d\NHI^\theta$ is given by
\<
 f_\theta(\NHI^\theta)d\NHI^\theta = \cos\theta f_\bot (\NHI^\bot) d\NHI^\bot ,
\>
thus we get
\< f_\theta(\NHI^\theta) = \cos^2\theta f_\bot (\NHI^\bot). \>
Finally, we average this distribution over all possible inclination angles to get the
column density distribution of randomly oriented galaxies:
\<
 f(\NHI) &=& \int_0^\frac{\pi}{2} d\theta \sin\theta f_\theta(\NHI) \nln &=&
\int_0^\frac{\pi}{2} d\theta \sin\theta \cos^2\theta f_\bot(\NHI/\cos\theta) \nln &=&
\frac{1}{\NHI^3} \int_0^{\NHI} d\NHI^\bot {\NHI^\bot}^2 f_\bot(\NHI^\bot) .
\label{eq:f_inc_avg} \>
While equation \eqref{eq:f_inc_avg} has appeared in the literature
\citep[e.g.][]{1995ApJ...454..698W,2006ApJ...652..981W}, previously it was derived for
column density distributions arising from galaxies with monotonically decreasing
radial profiles. Now we see that the result is more generic and applies to any two
dimensional structure, not simply disks. We would like to note that equation
\eqref{eq:f_inc_avg} implies that the column density distribution has a tail which
extends to arbitrarily large column densities. This is due to the relation between the
inclined and face-on column density, $\NHI^\theta = \NHI^\bot/\cos\theta$, which
implies that arbitrarily large column densities can be reached as we approach an
edge-on inclination. Of course, the real maximum is given by the maximum edge-on
column density. Therefore, even the randomly inclined column density distribution
naturally has a maximum column density. More precisely, this inclination model starts
to break down for large inclination angles when the lines of sight pierce uncorrelated
regions in the galaxy.

As we see from equation \eqref{eq:f_inc_avg}, if the face-on galaxy has a maximum
column density, or if the face-on galaxy column density distribution becomes steeper
than $\NHI^{-3}$ above some column density, then above this column density the column
density distribution will behave as a power-law with an exponent of $-3$. Note that
the high column density tail seen in \Figref{fig:zwaan_vs_prochaska_vs_noterdaeme} is
consistent with this $-3$ exponent.

With this idea in mind, we return to the plots we have above regarding the column
density distribution of the THINGS galaxies. As we can see in
\Figref{fig:THINGS_cdd_compare}, individual galaxies have steeper distributions than
the local column density distribution. This effect is further emphasized in
\Figref{fig:THINGS_turnover_vs_met} where we include inclination corrections and
determine the turnover column density by fitting a double power-law to the face-on
corrected column density distribution of each galaxy. Above the turnover column
density, the column density distributions of individual galaxies have slopes steeper
than $-3$. From equation \eqref{eq:f_inc_avg} we see that inclination averaging wipes
away the precise details of the column density distribution above this turnover column
density because the integral in equation \eqref{eq:f_inc_avg} is relatively
insensitive to steep column density distributions.

Now we can use this intuition to understand which systems are most important for the
column density distribution. Since inclination averaging wipes away information beyond
the turnover of an individual galaxy, the systems with column densities above this
turnover are not important for the global column density distribution, as measured for
a statistically large sample. These systems are necessarily small in size ($<$ few 100
pc) since they contribute relatively few lines of sight. Thus, we are arguing that the
small-scale features in column density maps are not important for the global column
density distribution whose high end tail is determined by projection effects of the
bulk of the gas in a galaxy.

\begin{figure}[t]
\includegraphics[width=8cm]{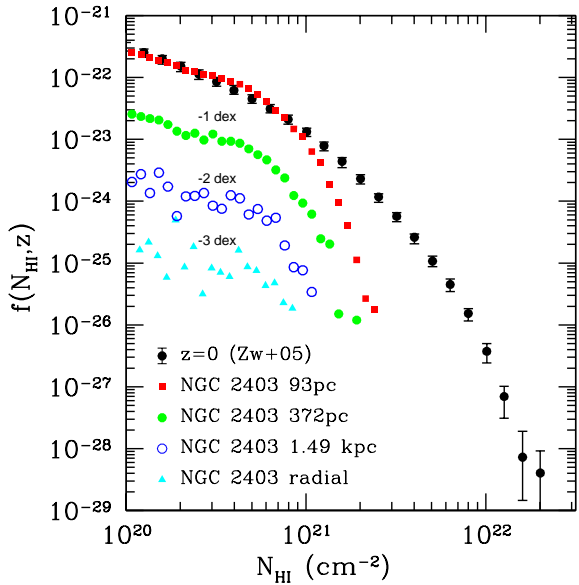}
\caption{Column density distribution of NGC 2403 corrected for inclination and computed
using maps of HI averaged on different scales. We compute the column
density distribution at different resolutions by first averaging the maps over patches
of the stated scale, and then computing the column density distribution from the resulting
maps. We note that the column density distributions are similar over a wide range of
scales, as expected due to the steep power spectrum of HI maps. We also show column
density distribution of this galaxy constructed from its radial surface density profile
(triangles), averaged in radial bins of 150 pc using the HI map at 93 pc resolution,
corrected for inclination, and using radii of the face-on projection.  The column density distribution
derived from the radial profile is similar to column density distribution derived from
a coarse grained map of the galaxy. For comparison purposes, we have offset each
subsequent column density distribution by 1 dex. In addition, we have included the
column density distribution of local galaxies as measured by \cite{2005MNRAS.364.1467Z},
listed as $z=0~~(\rm{Zw}+05)$.}
\label{fig:resolution_NGC_2403}
\end{figure}

To give further evidence to support this assertion, we consider the galaxy NGC 2403
from the THINGS sample. In \Figref{fig:resolution_NGC_2403} we display the face-on
corrected column density distribution of NGC 2403 computed using maps with pixels of
different size (indicated in the legend). In addition, we compute the column density
distribution as derived from the radial average surface density profile of NGC 2403.
As we can see, the column density distribution is quite insensitive to the averaging
scale of the map. As we average over progressively larger scales we simply lose the
relatively rare highest column density regions, while at lower column the distribution
is insensitive to the averaging scale. This insensitivity is due to the steep power
spectrum of HI maps, as noted in \Secref{sec:THINGS}. As a result, the turnover of the
column density distribution is almost identical over a wide range of scales. In
addition, the column density distribution derived from the radial profile has a
similar turnover. As a quantitative measure of this insensitivity to resolution, we
fit double power-laws to the distributions in \Figref{fig:resolution_NGC_2403} and
found that the turnover column density changed by 0.05 dex as we changed the
resolution from 93 pc to 372 pc, and by another 0.05 dex as we changed the resolution
from 372 pc to 1.49 kpc. This indicates that the scatter due to resolution effects in
\Figref{fig:THINGS_turnover_vs_met} should be quite small.

\begin{figure}[t]
\includegraphics[width=8cm]{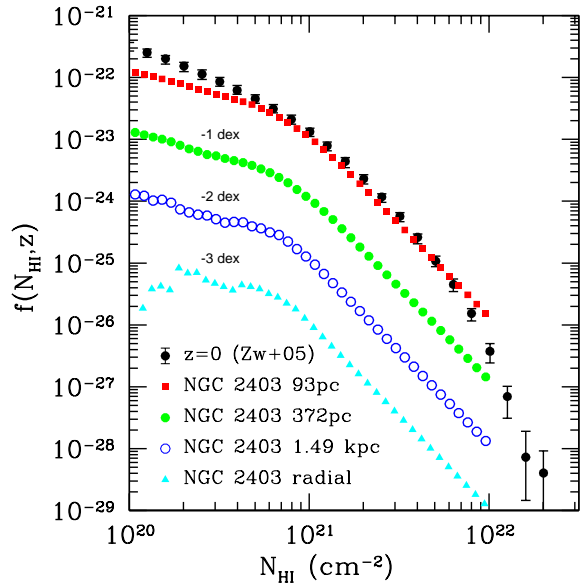}
\caption[]{Inclination averaged column density distribution of NGC 2403 computed at
various resolutions. To compute the inclination averaged column density distribution,
we take the face-on column density distributions from \figref{fig:resolution_NGC_2403},
and then use equation \eqref{eq:f_inc_avg}. The description of how we compute the
face-on column density distributions is given in \figref{fig:resolution_NGC_2403}. Once
again, we offset each distribution by 1 dex for comparison purposes. As we can see,
randomly orienting just NGC 2403 gives rise to a column density distribution which looks
remarkably similar to the local column density distribution from \cite{2005MNRAS.364.1467Z},
which is listed as $z=0~~(\rm{Zw}+05)$.
Furthermore, the inclination averaged column density distribution is very insensitive
to the scale at which it is computed, therefore the small-scale features are unimportant
for determining the global column density distribution. In light of the discussion after
equation \eqref{eq:f_inc_avg}, the turnover of the local column density distribution
above $\NHI \sim 10^{22} \pcms$ may also be due to the maximum column density of an
edge-on galaxy. Note that the dip in the radial column density distribution at small
column densities is due to the radial profile not extending to large enough radii and
hence small enough column densities. As a result, the face-on column density distribution
derived from the radial profile does not extend to low column densities so the inclination
averaged column density distribution will look artificially low for small column densities.}
\label{fig:inclination_NGC_2403}
\end{figure}

In \Figref{fig:inclination_NGC_2403} we show the inclination averaged column density
distribution computed using the face-on column density distribution from
\Figref{fig:resolution_NGC_2403} and equation \eqref{eq:f_inc_avg}. We see that the
inclination averaged column density distribution is very similar to the local column
density distribution from \cite{2005MNRAS.364.1467Z}. Note that we repeated this
procedure for the other galaxies in THINGS and got similar results. In addition, we
see that the inclination averaged column density distribution is even less sensitive
to the resolution than the face-on column density distribution. This is due to the
fact that equation \eqref{eq:f_inc_avg} is insensitive to the steepest parts of the
face-on column density distribution and hence insensitive to the small-scale
structures in the HI map. As a result, even the average radial surface density profile
of HI is sufficient to produce an inclination averaged distribution that is similar to
that obtained from the highest resolution HI map. Note that the high-$\NHI$ tail of
the column density distribution, $\NHI > 10^{22}\pcms$, can still be sensitive to the
small-scale structures in the map because even an edge-on galaxy will have some
maximum column density.

From \Figref{fig:THINGS_turnover_vs_met} we see that the turnover column density is
relatively similar for all galaxies in the THINGS sample and depends only weakly on
metallicity. Through inclination effects, this turnover column density within each
galaxy gives rise to the turnover in the column density distribution for a statistical
sample of galaxies. In light of this idea, we can now better understand the
universality of the column density distribution as seen in
\Figref{fig:zwaan_vs_prochaska_vs_noterdaeme}. The similarity in turnover of the
column density distribution at low and high redshift indicates that the galaxies at
low and high redshifts have similar maximum column densities of HI. We argue that this
universal maximum is not due to the HI-H$_2$ transition since this would imply a
dependence on metallicity which is not seen. The exercise described in this section
also indicates that this maximum column density does not arise in the small-scale
structure of the HI distribution, shaped, for example by turbulent cascades, but is
present in the large-scale structure of the HI disk. It is present, for example, in
the average radial HI surface density profile of disks. It thus appears to be related
to the global processes that set properties of galactic disks rather than the
small-scale dynamics of the ISM.

In principle, simulations of galaxy formation should also reproduce such universal
behavior if the physical processes shaping the surface density structure of galaxies
are modeled correctly. In the following section we will compare the observational
results discussed in this section to the results of cosmological galaxy formation
simulations.

%----------------------
\section{Simulations}
\label{sec:simulation_section}
%-----------------------------

In this section we will utilize our simulations in two ways. First, we will use the
observations described in the previous section as a stringent test of our simulations.
Second, we will use the simulations as a testing ground to investigate the effects of
various physical mechanisms on the column density distribution. This will teach us
which of these physical mechanisms are most relevant for the turnover.

In this paper we use several sets of related simulations performed with the Adaptive
Refinement Tree (ART) code
\citep{1999PhDT........25K,2002ApJ...571..563K,2008ApJ...672...19R}, which uses
adaptive mesh refinement for both gas and dark matter, thus achieving a large dynamic
range in spatial scale. These simulations follow one or more regions of interest,
typically selected as Lagrangian regions of five virial radii around typical dark
matter halos. Regions of interest are embedded into a coarsely resolved cosmological
cubic volume with periodic boundary conditions.  In all simulations used here the
maximum spatial resolution inside regions of interest is about $260$ comoving pc ($65$
pc at $z=3$) which is the result of 9 levels of adaptive mesh refinement. The highest
dark matter mass resolution is $1.3 \times 10^{6} M_\odot$ and the baryonic mass
resolution varies from $\sim 10^3 M_\odot$ to $\sim 10^6 M_\odot$ depending on cell
size and density.

All our simulations include three-dimensional radiative transfer of UV and ionizing
radiation from stars formed during the simulation, as well as the cosmic background.
This is done with the OTVET approximation \citep{2001NewA....6..437G}. Radiative
transfer is especially important for the column density distribution since it
correctly models the local radiation flux which can ionize neutral hydrogen regions.
Finally, these simulations include the non-equilibrium chemical network of hydrogen
and helium described in \cite{2011ApJ...728...88G}. This network includes molecular
hydrogen and, in particular, the formation of H$_2$ in both the primordial phase and
on dust grains. We note that the H$_2$ model in \cite{2011ApJ...728...88G} is
calibrated against observations of the Milky Way, LMC, and SMC. Therefore, we expect
the model to be valid over a range of metallicities and radiation fields which extend
to at least the values characterized by the SMC.

Our fiducial run is a fully cosmological simulation with a box size of $6h^{-1}$
comoving Mpc and a single region of interest around a Milky-Way type galaxy \citep[the
simulation called ``cosmo II'' in][]{2011ApJ...728...88G}. The computational box
outside the region of interest is covered with a uniform $64^3$ grid. The cosmological
parameters used in the fiducial run are similar to the WMAP1 parameters: $\Omega_M =
0.3, \Omega_B = 0.046, \sigma_8 = 0.9,$ and $h=0.7$.

In addition to the fiducial run, we use a larger run with a $25h^{-1}$ comoving Mpc
box and $256^3$ root grid, focusing on 5 Lagrangian regions around halos with masses
between $10^{12}$ and $10^{13} M_\odot$, with exactly the same physics and mass and
spatial resolution as in the fiducial run. We use this simulation to test the robustness of
our results and for better statistics of halo based properties.

We also used the ``fixed ISM" runs of \citet{2011ApJ...728...88G}, in which the
dust-to-gas ratio and the overall normalization of the radiation field at $1000 {\AA}$
were fixed to constant values. These are useful when we want to compare particular
galaxies to models with exactly the same dust-to-gas ratio and interstellar UV field.
For example, as a model for the ISM of the SMC we can use a ``fixed ISM" simulation
with a dust-to-gas ratio corresponding to a metallicity of $Z \approx 0.1 Z_\odot$ and
a radiation field which is $10-100$ times larger than that in the Milky Way.

\subsection{Computing the Simulated Column Density Distribution}

In order to compare our simulations to the observations in
\Secref{sec:observational_tension}, we need to compute the column density
distribution. The procedure is similar to the one discussed for the THINGS galaxies in
\Secref{sec:THINGS}.

Since the DLAs are rare in our simulation, we expect to find at most one DLA along a
random line of sight. Therefore, we project our simulated box separately along each
axis to get the projected column densities for the box. On the boundary we use a grid
whose cell size is the same as the smallest cell in our simulation. For the $6$
comoving $h^{-1}$ Mpc box we project onto a grid consisting of $(64 \times 2^9)^2$
pixels which corresponds to a resolution of $65$ pc at $z=3$. Given this projected
map, we can compute the column density distribution using equation
\eqref{eq:discretef}. To improve our statistics, we average the column density
distribution over the three cartesian projections. We have verified that the column
density distribution has converged at this resolution for column densities below $\NHI
\approx 10^{23} \pcms$.

In addition, when we project the box onto the boundary grid as described above, we
also save the location of the highest HI density cell along each hypothetical line of
sight. This is useful since we can subsequently project the HI density along a shorter
segment which is centered on the maximum HI density cell. We have verified that if we
integrate along a $27$ comoving kpc segment about each of these maxima and compute the
column density distribution, we have converged to the full-box column density
distribution above $\NHI = 10^{19}\pcms$. This is useful when we post-process the
simulation since then we can save computational time by integrating along these
relatively short lines of sight.

Note that the column density distributions considered in this study were extracted
from Lagrangian regions around progenitors of massive galaxies and therefore they may
not be fully representative of the mean distribution in a large random volume of the
universe. However, we focus on qualitative features of the column density distribution
and do not attempt to make detailed quantitative comparisons. Given that the
discrepancies we identify can be traced to the internal density distribution in
galaxies of a wide range of masses, we believe our results and conclusions are
representative and generic.

\subsection{Comparing the Simulated Column Density Distribution to Observations}

\begin{figure}[t]
\includegraphics[width=8cm]{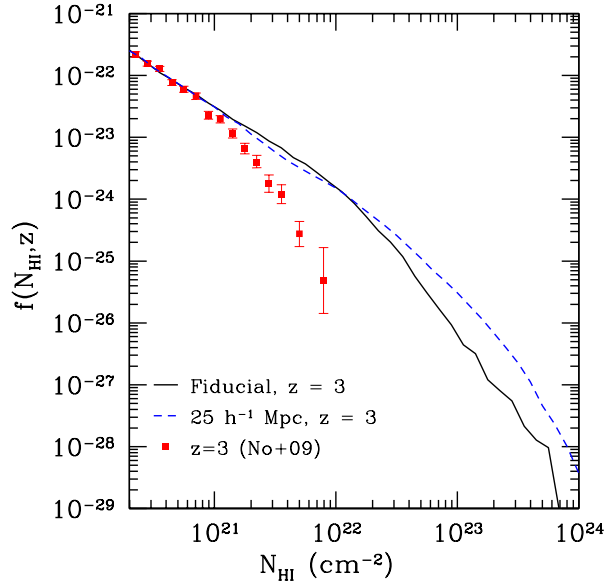}
\caption{Comparison of the fiducial run with observations at $z=3$. The squares, $z=3~~(\rm{No}+09)$, are from
\cite{2009A&A...505.1087N}. Clearly, we overpredict the frequency of high column density DLAs.}
\label{fig:fiducial_vs_noterdaeme}
\end{figure}

A comparison of the observed column density distribution to simulation results is
shown in \Figref{fig:fiducial_vs_noterdaeme} for both the $6$ and $25 h^{-1}$ comoving
Mpc boxes. Note that we have normalized the column density distributions to match the
observed column density distribution at $\NHI = 10^{20.3} \pcms$.

From \Figref{fig:fiducial_vs_noterdaeme} we see that the simulated column density
distribution turns over at a higher column density than the observed column density
distribution. Since we have a large simulation volume which includes many galaxies,
this result has already taken into account the effect of random orientations which we
described in \Secref{sec:inclination}. Furthermore, since our simulations have
metallicities and radiation fields which are consistent with Lyman Break galaxies over
a variety of redshifts \citep[see Fig. 2 in][]{2010ApJ...714..287G} and our HI-H2
model is consistent with observations, we should be correctly reproducing the effect
of the HI-H2 transition. The fact that our HI column density distribution is turning
over at such a high column density indicates that our galaxies have face-on column
density distributions which extend to high column density. This indicates that we are
not correctly modeling the surface density distribution of the individual galaxies in
the simulation. Below we will investigate how this can be remedied and what it tells
us about the physics which shapes the column density distribution.

\subsection{Effect of Metallicity and Radiation Field on the Column Density
Distribution}

\begin{figure}[t]
\includegraphics[width=8cm]{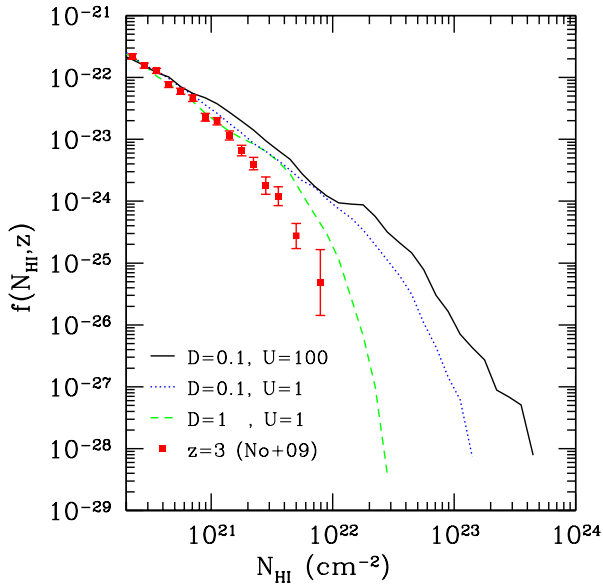}
\caption{Comparison of column density distributions for various ISM conditions. The squares, $z=3~~(\rm{No}+09)$, are from
\cite{2009A&A...505.1087N}. The dust-to-gas ratio and the radiation fields
are in solar and Milky Way units respectively. As we see, for sufficiently high metallicity, we can reproduce the
observed column density distribution.}
\label{fig:diff_ISM_comparison}
\end{figure}

A large part of our argument in \Secref{sec:observation_section} was based on the
effect of metallicity and UV field on the HI-H$_2$ transition and the corresponding
effect this would have on the column density distribution. Since our simulations model
molecular hydrogen, we can test this directly in the simulations using the fixed-ISM
runs. We considered SMC-like ISM parameters: $Z = 0.1 Z_\odot$ and a UV radiation
field, denoted by $U$, which is $100$ times stronger than the fields in the Milky Way,
$U = 100 U_{MW}$; and Milky Way-type ISM parameters: $Z = Z_\odot$ and $U = U_{MW}$.
Finally, we consider an intermediate model with $Z = 0.1 Z_\odot$ and $U = U_{MW}$.
The column density distributions of these three model ISMs are presented in
\Figref{fig:diff_ISM_comparison}.

As we can see from \Figref{fig:diff_ISM_comparison}, an increase in the metallicity
shifts the turnover to lower column density. This is expected due to the metallicity
dependence of the HI-H$_2$ transition. As we argued above, the observed turnover at
$\NHI\sim 10^{21} \pcms$ does not depend strongly on metallicity
(\figref{fig:zwaan_vs_prochaska_vs_noterdaeme} and \figref{fig:THINGS_cdd_compare})
and thus is not due to HI-H$_2$ transition.  As a check that the difference in
\Figref{fig:diff_ISM_comparison} is indeed due to the HI-H$_2$ transition and not some
process which is blowing all the high density gas away, we can compare HI column
density distribution to the neutral hydrogen column density distribution, which
includes both atomic and molecular gas. Such distributions are very similar for all of
the metallicity and radiation field configurations in
\Figref{fig:diff_ISM_comparison}, which indicates that the trend shown is indeed due
to the metallicity dependence of the HI-H$_2$ transition.

\subsection{Systems Responsible for the High Column Density Discrepancy}

\begin{figure}[t]
\includegraphics[width=8cm]{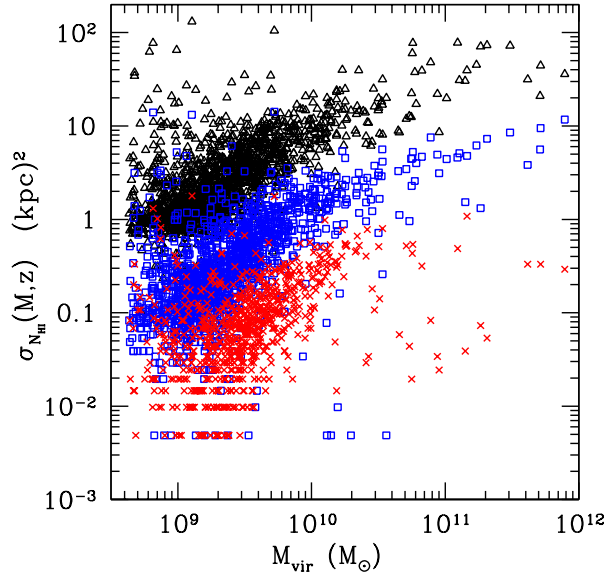}
\caption{Proper cross-section versus the virial mass of the nearest primary halo. This
data is from the $25 h^{-1}$ comoving Mpc simulation to get better halo statistics.
The triangles represent the cross-sections of halos above the DLA threshold of $\NHI = 2\times 10^{20}\pcms$.
The squares represent the cross-sections of halos above $\NHI = 10^{22}\pcms$. The crosses
represent systems above $\NHI = 10^{23}\pcms$.}
\label{fig:x_section_vs_mvir}
\end{figure}

In order to understand which physical process are missing or incorrectly modeled in
our simulations, it is helpful to know which systems are contributing the most to the
high column density lines of sight. This is accomplished by associating the HI systems
with their host halo and determining which halos give rise to the high column density
systems. First, we consider the cross-section above a fixed column density versus the
virial mass which is plotted in \Figref{fig:x_section_vs_mvir}. In terms of the
quantities in \Secref{sec:background_cdd}, the cross-section for a halo of mass $M$
above a HI column density of $\NHI$ is given by
\< \sigma_{\NHI}(M,z) = \int_{\NHI}^\infty \frac{\partial \sigma(M,\NHI',z)}{\partial
\NHI'} d\NHI' .\>
From \Figref{fig:x_section_vs_mvir} we can see how the cross-section of a galaxy above
a fixed column density increases with increasing halo mass. Although the more massive halos have
larger cross-sections, we must weight the cross-sections by the number density of
their associated halos in order to know the relative importance of each galaxy type.

\begin{figure}[t]
\includegraphics[width=8cm]{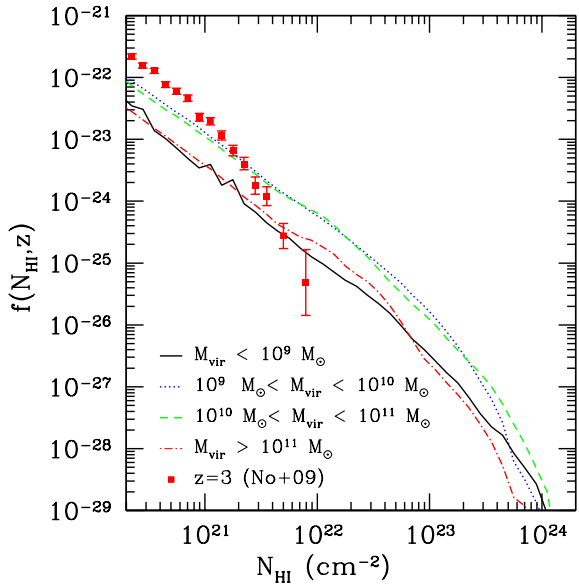}
\caption{Contribution to the column density distribution from different
ranges of halo masses. The lines are obtained from the $25 h^{-1}$ comoving Mpc simulation
at $z=3$. The squares, $z=3~~(\rm{No}+09)$, are observations from
\cite{2009A&A...505.1087N}. Note that the column density distributions here are
normalized such that their sum matches the observed distribution at $\NHI = 10^{20.3}\pcms$. }
\label{fig:cdd_halo_mass_cuts}
\end{figure}

One way to describe the importance of each mass range is to compute the column density
distribution of systems associated with each mass range. This is shown in
\Figref{fig:cdd_halo_mass_cuts}. As we can see, the shape of the column density
distribution is fairly similar for all mass ranges. While the dwarf galaxies, with
halo masses in the range $10^{9} M_\odot < M < 10^{11} M_\odot$, give the largest
contribution at all column densities, none of the mass ranges have column density
distributions which are consistent with observations. Therefore, we see that we need
to include physics which is efficient at removing high density gas from both dwarf
galaxies and Milky Way-like galaxies.

\subsection{Additional Physical Mechanisms which can Affect the Column Density Distribution}

Now that we understand which systems are giving rise to the discrepancy, we can try to
understand what additional physics we need to correctly model these systems.

Looking back to equation \eqref{eq:intCDD}, we recall that we must correctly model the
differential cross-section, or equivalently the surface density distribution, of each
galaxy to get the correct column density distribution. We are already modeling the
HI-H$_2$ transition using the model of \cite{2011ApJ...728...88G} and observational
evidence indicates that this transition does not play an important role for these
systems. As a check we have also post-processed our gas with the H$_2$ model of
\cite{2008ApJ...689..865K,2009ApJ...693..216K,2010ApJ...709..308M}, which leads to
only a small decrease in the column density distribution function at large $\NHI$. The
small difference between our model and the model of Krumholz et al is not surprising,
given that \cite{2011ApJ...729...36K} showed that these two H$_2$ models agree well
for metallicities greater than $Z/Z_\odot \sim 0.1$.

\subsubsection{Sub-grid Models for Turbulence and Self-gravitation}

While the observed column density distribution at high redshift is probing the gas at
scales significantly smaller than $1$ pc, the highest resolution of our simulations is
only $65$ pc at $z=3$. Therefore it is possible that the discrepancy between observed
and simulated column density distributions is due to different column density
structure on small scales. Indeed, quasar lines of sight probe the column density
distribution on sub-parsec scales, while the smallest resolved scales in simulations
are $\sim 100$~pc. It is not guaranteed {\it a priori} that column density
distributions measured at such different scales should match. As an extreme example,
consider a toy model in which the HI gas on small scales is in tiny $\sim 1$ pc size
clouds. For a sparse scattering of clouds measured at a resolution of $1$ pc or
smaller, most lines of sight would pierce no clouds and a few would pierce the cloud
giving rise to two peaks in the column density distribution. In contrast, at a
significantly coarser resolution, say $100$ pc, the column density distribution would
depend on the spatial distribution of these clouds. If they are uniformly spread out
then the column density distribution would have a single peak at the average column
density with a width given by the central limit theorem. However, in general,
different regions could have different column densities when averaged over $100$ pc,
giving rise to a very broad column density distribution. In principle, the column
density distribution measured at different scales can look very different.

We can attempt to model the effect of a physically plausible small-scale density
distribution using the results of turbulence simulations on GMC scales. Turbulence
leads to compression and rarefaction of the gas, which can affect the column density
of the gas and generically predicts a log-normal distribution for densities. While we
do not know the exact nature of DLA systems at high redshift, observations of the
local universe indicate that the highest column density systems are in structures like
giant molecular clouds (GMCs). The gas density PDF in such clouds has been studied and
is known to be log-normal \citep[see][for a
review]{2007ARA&A..45..565M,2007ApJ...665..416K}. Interestingly, the column density
PDF also obeys a log-normal distribution in GMCs
\citep{2001ApJ...546..980O,2009ApJ...692...91G}. In \cite{2001ApJ...546..980O} the
authors simulated GMCs using magnetohydrodynamics and found that the column densities
obeyed a log-normal PDF which indicated that the gas density was correlated along
lines of sight due to coherent structures on larger scales. This was followed up by
observational results in \cite{2009ApJ...692...91G} where the authors observed the
column density distribution of nearby GMCs. The observed column densities satisfied a
log-normal PDF with similar variance to that of \cite{2001ApJ...546..980O}.

These studies found that the area-weighted column density PDF of a cloud with a mean
column density of $\overline{N}$ is given by
\< P(y) = \frac{1}{\sqrt{2\pi \sigma^2}} \exp\Big(-\frac{1}{2\sigma^2}(y+\mu)^2\Big),
\>
where $y=\ln(N/\overline{N})$ and $\mu = \sigma^2/2$, as required for normalization.
\citet{2001ApJ...546..980O} and \citet{2009ApJ...692...91G} found that the log-normal
distribution is a good fit to the simulated and observed distributions with $\sigma
\in (0.25,0.55)$ depending on the conditions in the gas, ie mach number and magnetic
field. In comparison to these GMCs, the DLAs consist of mostly atomic hydrogen, which
can be expected to have a smoother spatial distribution. Therefore, the column density
PDFs described above will serve as an overestimate of the variance in DLAs. In light
of this, we have tested a large range of parameters, $\sigma \in (0.1,5)$, to see if
there is any large effect on the column density distribution.

\begin{figure}[t]
\includegraphics[width=8cm]{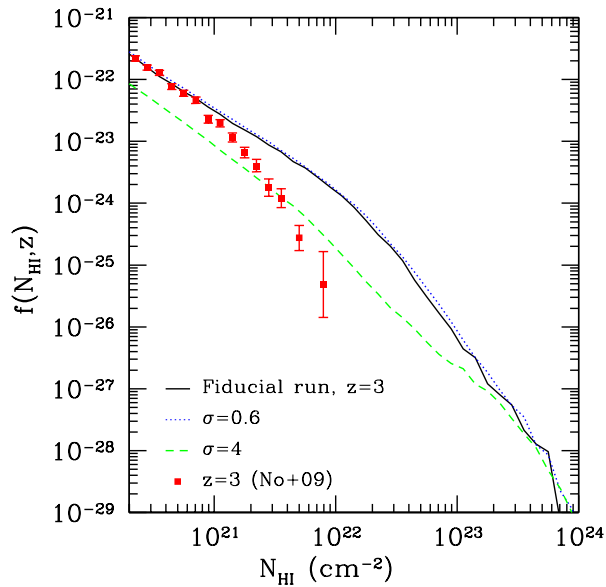}
\caption{Effect of the log-normal PDF of ISM turbulence on the
column density distribution. The squares, $z=3~~(\rm{No}+09)$, are observations from
\cite{2009A&A...505.1087N}. The highest observed values are $\sigma \sim 0.6$. We
include the extreme case of $\sigma = 4$ to show that even an extreme log-normal PDF
will not remedy the discrepancy. Note that we have normalized the $\sigma = 0.6$ and
$\sigma = 4$ runs by the same amount for the sake of comparison.}
\label{fig:log-normal_pdfs}
\end{figure}

We implement the log-normal PDF on a cell-by-cell basis. In each cell, along each line
of sight, we take the HI column density and multiply it by a realization of the
log-normal PDF. This procedure rarefies and compresses adjacent cells independently
which is justified since our minimum cell size is $65$ pc at $z=3$, larger than a
typical GMC. Since this procedure changes the column density and hence the shielding
properties of the gas, we also re-process the resulting compressed or rarefied gas
with fits to the H$_2$ models in \cite{2011ApJ...728...88G}. The processed column
density distributions are presented in \Figref{fig:log-normal_pdfs}. In addition, we
experimented with applying the realization only if the individual cell has a
sufficiently large $\NHI$, but the difference is negligible above this column density
threshold.

As we can see from \Figref{fig:log-normal_pdfs}, there is very little effect for
reasonable values of the variance. However, for larger values we see that the normalization of the column
density distribution is decreased. This follows since the
log-normal PDF is not symmetric and rarefies more systems than it compresses. We have
also checked that this post-processing model of turbulence can be implemented by
taking the convolution in log-space of the HI column density distribution with a
log-normal PDF.

Another important subgrid mechanism that can affect these high column density systems
is self-gravitation. Observations of molecular clouds and simulations have indicated
that self-gravitation gives rise to a power-law tail in the column density PDF at high
column densities \citep{2011ApJ...727L..20K}. We can include this effect by
implementing the same procedure as for the log-normal PDF but also including a
power-law tail.

Since we do not know the exact power-law behavior that will be present in high
redshift DLAs we once again test a variety of parameters. For realistic power-law
tails similar to the ones in \citet{2011ApJ...727L..20K}, we find no significant
change in the column density distribution. This is expected from cross-sectional
arguments since the regions in which the power-law behavior develops, GMCs, will be
very small and hence very rare in a cross-section weighted sample.

As a final comment, we note that the results of the log-normal PDF model described
above are consistent with an independent, observation-based argument mentioned in
\Secref{sec:THINGS}. There we noted that observations of nearby galaxies show that the
HI column density power spectra can be described by a steep power-law over a wide
range of scales, ranging from $\sim 0.1$ pc up to $\sim 10$ kpc
\citep[][etc.]{1999MNRAS.302..417S,2001ApJ...548..749E,2003A&A...411..109M,2009MNRAS.398..887D},
with slopes of the power spectra ranging from $\sim -1.5$ to $\sim -3.5$. Such steep
power-law behavior indicates that the HI maps do not have large-amplitude fluctuations
on small scales and hence the column density distribution cannot be a strong function
of averaging scale. This argument hinges on observations of local galaxies. However,
if the origin of steep power-law spectra is due to some universal process, such as
turbulence, it is reasonable to expect that ISM giving rise to DLAs will have
similarly steep power spectra.

\subsubsection{Toy Model for Feedback Effects}

Since DLAs are associated with the ISM of galaxies, they can be greatly affected by
the physical processes which occur in the central regions of these galaxies where the
star formation rate is highest and effects of feedback can be expected to be
strongest. Due to high densities and star formation rates, the centers of these
galaxies will be affected by feedback effects like supernovae and radiation pressure.
Some of these effects have been included in recent works
\citep[see][]{2011ApJ...737L..37A,fumagalli_etal11}. While we have included the
effects of supernovae, this feedback scheme is inefficient in driving outflows and
does not change the distribution of gas appreciably.

To explore the impact of a much stronger feedback, we consider a simple toy model.
Since feedback mechanisms are most relevant for the high densities in the inner parts
of galaxies, we can set an upper limit of their effect by removing the inner parts of
these disks when we calculate the column density distribution. If this overestimate of
the feedback effect is compatible with the observed column density distribution, then
it is plausible that feedback can bring our simulation into agreement with
observations.

\begin{figure}[t]
\includegraphics[width=8cm]{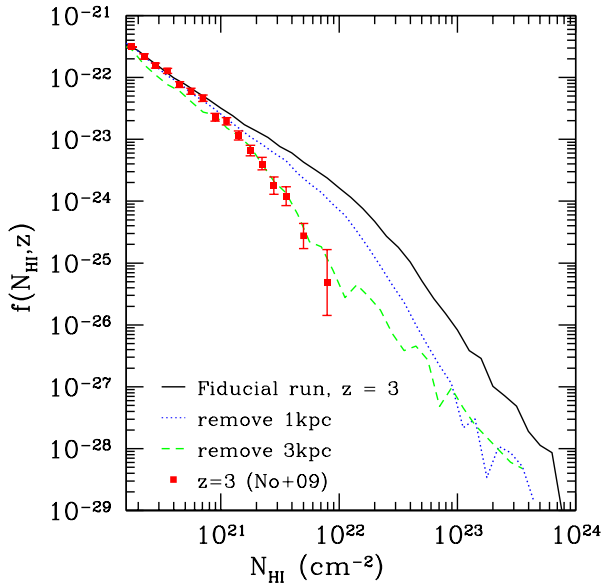}
\caption{Effect on column density distribution of removing different sized regions around
the centers of halos. The squares, $z=3~~(\rm{No}+09)$, are observations from
\cite{2009A&A...505.1087N}. We see that removing the gas in the inner $3$ kpc of halos brings
the column density distribution into agreement with observations. }
\label{fig:remove_inner_comparison}
\end{figure}

\Figref{fig:remove_inner_comparison} shows the column density distribution with
regions of radius $1$ and $3$ kpc from the parent halo center removed. We see that
removing a region of radius $3$ kpc is sufficient to bring the simulations into
agreement with observations. It remains to be seen if realistic star formation and
feedback models can account for such a substantial removal of gas from the central
regions of high-redshift galaxies.

\section{Comparison to recent simulations}
\label{sec:comparison_section}

A number of recent studies have presented column density distributions from
simulations and compared them to observations
\citep[e.g.][]{2006ApJ...645...55R,2008MNRAS.390.1349P,2010arXiv1010.5014C,2010ApJ...725L.219N,2011ApJ...737L..37A,fumagalli_etal11}.
Some of the simulations in these studies have successfully reproduced the column
density distribution while others were discrepant. Since the observed column density
distribution is a sub-parsec scale probe of atomic hydrogen at high redshift, it
provides a stringent test of simulations. In fact, since the resolutions found in
cosmological simulations are at best hundreds of parsecs, it is surprising that any
simulation can match observations without delving into the sub-grid physics. In this
section we will try and understand the reasons behind the success and failures of
these simulations. Such insight will help us understand what additional physical
processes are needed in future simulations to obtain the observed column density
distribution.

\cite{2006ApJ...645...55R} used an adaptive mesh refinement simulations with a set of
physical processes similar to our simulations. Their simulations include detailed gas
chemistry for all species of hydrogen and helium, radiative transfer above the lyman
limit, UV background, but no star formation and no H$_2$ model for HI-H$_2$
transition. They simulated a variety of box sizes ranging from $2-8 h^{-1}$ comoving
Mpc with resolution at $z=3$ varying from 93 pc for the smallest box, to 1.5 kpc for
the largest box. In addition, they ran their $8h^{-1}$ comoving Mpc box with a variety
of resolutions from 190 pc to 1.5 kpc. \cite{2006ApJ...645...55R} found a power-law
column density distribution in their simulations that extends to $\NHI\sim
10^{22.5}\pcms$ without any sign of a turnover. We find a similar behavior which
extends to even higher column densities due to the higher resolution of our
simulation.

\citet{2010arXiv1010.5014C} used an adaptive mesh refinement code with a zoom-in
simulation on a $120 h^{-1}$ comoving Mpc box onto a cluster and a void of sizes
approximately $20 h^{-1}$ comoving Mpc and $30 h^{-1}$ comoving Mpc respectively. The
grid resolution is $460 h^{-1}$ pc. The simulations include a Haardt/Madau UV
background, neutral hydrogen self-shielding, metallicity dependent radiative cooling,
star formation, supernova feedback, and a molecular hydrogen model based on a solar
metallicity HI-H$_2$ transition. They find a column density distribution at $z=3$
which is similar to observations. The key difference of our simulations is that we
include a model for HI-H$_2$ transition, which takes into account the dependence on
metallicity. The model used by \citet{2010arXiv1010.5014C}, on the other hand, is
based on observations of the HI-H$_2$ transition in the Milky Way. As a check of the
effect of their H$_2$ model, we have reprocessed our neutral hydrogen with the H$_2$
model used by  \citet{2010arXiv1010.5014C} and found that we also get a similar
agreement with observations. However, this agreement is artificial because a solar
metallicity model for the HI-H$_2$ transition is not applicable to $z\sim 3$ DLAs,
which have metallicities $Z\lesssim 0.1Z_{\odot}$. Therefore, we can conclude that the
HI column density distribution in simulations of \citet{2010arXiv1010.5014C} actually
disagrees with observations.

\citet{2011ApJ...737L..37A} used the set of the OWLS simulations, run with Gadget SPH
code with subgrid models for star formation, chemodynamics, galactic winds, cooling in
the presence of a uniform UV background, and a Milky Way-type H$_2$ model. They
consider column density distribution in a box of size $25 h^{-1}$ comoving Mpc with a
gravitational softening length of $2.79$ comoving kpc. With this set of physical
processes, \citet{2011ApJ...737L..37A} were able to approximately reproduce the column
density distribution at $z=3$ across many decades in column density, $\NHI \sim
10^{12.5} - 10^{22} \pcms$. Notably, they get the correct turnover in the column
density distribution at $\NHI \approx 10^{21.5} \pcms$. One potential caveat to this
result is the relatively low resolution of the OWLS simulations. However, we have
re-run our simulations at a similar resolution, a minimum cell size of $520$ pc at
$z=3$, and found that such a decrease in resolution does not lead to a better
agreement with the observed column density distribution. In addition, the OWLS
simulations use a star formation prescription which relies on a rigid equation of
state which can, in principle, prevent the formation of high density regions within
the ISM. Interestingly, \citet{2011ApJ...737L..37A} find that the HI-H$_2$ transition
affects the HI column density distribution only at $\NHI\gtrsim 10^{21.7}\pcms$, which
is consistent with our findings. However, since their H$_2$ model is based on an
empirical calibration of the molecular fraction by \citet{blitz_rosolowsky06}, which
used local galaxies, the resulting calibration may thus not be applicable to
low-metallicity ($Z/Z_{\odot}\lesssim 0.1$), high-redshift DLA systems
\cite{2010ApJ...722..919F}. Indeed, although the galaxy sample studied by
\citet{blitz_rosolowsky06} included two galaxies with metallicities of $\sim
0.2Z_{\odot}$, the majority of galaxies in their sample have metallicities close to
solar. As we noted before, strong dependence of the HI-H$_2$ transition on metallicity
is both expected theoretically and is actually observed in measurements of molecular
fraction along stellar lines of sight in the Milky Way, LMC, and SMC
\citep{2002ApJ...566..857T,2006ApJ...636..891G,ism:wthk08}.

In \citet{fumagalli_etal11} the authors utilize an old version of the ART code, which
differs from the version we use in its implementation of star formation, cooling, and,
stellar feedback. The resolution of their simulations is $\approx 35-70$ pc depending
on redshift. \citet{fumagalli_etal11} show that their simulations successfully
reproduce the turnover in the column density distribution function. One potential
reason for this match is their aggressive star formation prescription and supernova
feedback. The star formation efficiency in their simulations is 20-50 times higher
than the efficiency we use and the efficiency that is estimated for real molecular
clouds. Likewise, their feedback includes processes that overestimate the actual
amount of energy that should be injected into the ISM. Their aggressive star formation
thus likely disrupts regions of high HI column density gas and may be responsible for
their agreement with the observed column density distribution function. Although we
consider the star formation and feedback model employed in these simulations extreme,
the results do indicate that the discrepancy in column density distribution between
observations and our simulations is likely related to the inefficient feedback of our
simulations. In their appendix, these authors show that inclusion of the H$_2$ model
of \cite{2008ApJ...689..865K,2009ApJ...693..216K,2010ApJ...709..308M} is only
important above $\NHI \sim 10^{22} \pcms$, once again in agreement with our argument
that the HI-H$_2$ transition is not relevant for the turnover in the column density
distribution.

As we can see from this section, several different mechanisms can lead to the correct
column density distribution. Even if we doubt the physical applicability of these
mechanisms, we can still learn a lot from them. Clearly, star formation and feedback
have a large effect on the column density distribution, while the HI-H$_2$ transition
is unimportant for the turnover at $\NHI \sim 10^{21}\pcms$. Therefore, future efforts
to correctly model the column density distribution will require correctly modeling
star formation and feedback and, more generally, the density structure of turbulent
galactic disks. We plan to explore physical processes that affect the density
structure of the ISM using high-resolution simulation models of the ISM in a future
study.

\section{Conclusions and Future Directions}
\label{sec:conclusion}

In this study, we have argued that there are two features in the column density
distribution at high HI column density systems. First, there is a turnover at $\NHI
\sim 10^{21} \pcms$ which is present in both $z=0$ and high-redshift HI column density
distributions. Second, at low redshifts there is also a lack of high column density
systems above $\NHI \sim 10^{22} \pcms$. This second turnover can be plausibly
explained by   the HI-H$_2$ transition. At this point, however, it is not clear
whether a similar high-column density turnover exists in the high-$z$ distribution as
probing this regime would require orders of magnitude larger absorber samples. In this
study we have focused on the universal turnover at $\NHI \sim 10^{21} \pcms$.

In \Secref{sec:observational_tension} we argued that the HI-H$_2$ transition does not
contribute significantly to the turnover. This was supported by a comparison of
observational evidence at both high and low redshift which showed an identical
turnover. This is in contrast to what is expected from a HI-H$_2$ induced turnover
which would be strongly metallicity dependent. Furthermore, we showed that even in the
local universe, the column density distributions {\it within} nearby galaxies exhibit
a universal shape similar to that of the overall column density distribution and show
little dependence on metallicity. Altogether, this observational evidence led us to
conclude that the turnover is due to some universal properties of HI distribution in galaxies,
presumably shaped by global processes that set overall structure of gaseous disks during their formation.

In the following section, \Secref{sec:inclination}, we revisited the idea of deriving the
column density distribution from randomly oriented disks which was originally considered
in \cite{1988A&A...202L...9M}, \citet{fall_pei93}, \citet{1995ApJ...454..698W}. We
extended this idea to randomly oriented two dimensional structures in equation
\eqref{eq:f_inc_avg}. This extension allows us to take the face-on column density
distribution of a single galaxy and then average it over inclination angles to produce a
statistical column density distribution. As we saw from
\Figref{fig:inclination_NGC_2403}, this inclination averaged column density distribution
is remarkably similar to the local column density distribution as measured in
\cite{2005MNRAS.364.1467Z}. Furthermore, we showed that this inclination averaged column
density distribution is insensitive to the averaging scale of the HI map, indicating that
the local column density distribution itself is fairly insensitive to small-scale
features. This implies that the turnover at $\NHI \sim 10^{21} \pcms$ is not due to the
small-scale features in a galaxy, but rather due to the large-scale structure of the HI
distribution or, equivalently, the radial profile of the galaxy.

Next, we turned to comparisons of the HI column density distribution in our
simulations and in observations, which showed that our simulations severely
overpredict the frequency of high column density systems, especially above $\NHI
\approx 10^{22}\pcms$. Since the column density distribution depends on a variety of
physical processes, our failure to match the observed distribution teaches us about
the shortcomings of our simulations. We used the simulations to explicitly show how a
HI-H$_2$ induced turnover depends on metallicity. We also showed that dwarf galaxies
at $z\sim 3$ provide a dominant contribution to the column density distribution, but
Milky Way-sized galaxies also contributed some high column density sight lines. We
showed that the discrepancy is not alleviated by increasing resolution.

We then checked the effect of two important pieces of sub-grid physics on the column
density distribution: {\it small-scale} ISM turbulence and gravitational collapse
within GMCs. Both of these were included through their known effect on the column
density PDF. We found that for physically realistic parameters, they have a negligible
effect. In addition, we showed that simulations can be brought in agreement with
observations if a region of radius $3$ kpc around centers of all galaxies is removed.
This exercise shows that the discrepancy is due to excessively high densities in the
centers of our simulations. Presumably this excess gas should be removed by stellar
feedback. However,  it remains to be seen if realistic models of stellar feedback can
remove such a large amount of gas.

Finally, we compared our results to the results of several recent papers which
presented simulation predictions for the HI column density distribution. Although some
of the simulations in these studies were successful in reproducing the high column
density part of the distribution, we believe that this success may partially be due to
the extreme feedback models employed. Despite this, their success indicates the
sensitivity of the column density distribution to these mechanisms and indicates that
the high-$\NHI$ tail of the DLA column density distribution is a sensitive probe of
both gas dynamics in high-redshift disks and effects of stellar feedback. Future work
can therefore fruitfully focus on detailed modeling of these aspects of forming
high-$z$ galaxies.

Shortly after the submission of this manuscript, \cite{2012ApJ...749...87B} considered
the effect of 21cm HI self-absorption on the column density distribution. He claims
that the column density distribution is sensitive to both resolution, in contradiction
with our argument in \Secref{sec:THINGS} and \Figref{fig:resolution_NGC_2403}, and
opacity corrections. In Figure 6 of \cite{2012ApJ...749...87B} he compares the HI
column density distribution of a large sample of nearby galaxies
\citep{2005MNRAS.364.1467Z}, which made use of $\sim 1.5$ kpc to $\sim 3$ kpc-scale
maps, to the column density distribution of a weighted combination of three local
galaxies (M31, M33, and LMC) corrected for self-absorption, which made use of $\sim
100$ pc-scale maps. While these column density distributions have a significant
difference below $\NHI=10^{21} \pcms$, opacity corrections are not important at such
low column densities. This indicates that the difference is due to the fact that M31,
M33, and LMC do not form a representative sample. Above $\NHI=10^{21} \pcms$ the
differences are small but since the sample used to make the opacity corrected column
density distribution is not representative, it is not clear how significant this
difference is.

\acknowledgements

We would like to thank Art Wolfe for discussions, careful reading of a preliminary
draft of this paper and comments that improved the content and presentation. We would
also like to thank Sam Leitner, Robert Feldmann, and Craig Booth for useful
discussions. Finally, we thank the anonymous referee for their helpful and
constructive comments. This work was supported in part by the DOE at Fermilab, by the
NSF grant AST-0708154, by the NASA grant NNX-09AJ54G, and by the Kavli Institute for
Cosmological Physics at the University of Chicago through the NSF grant PHY-0551142
and PHY-1125897 and an endowment from the Kavli Foundation. The simulations used in
this work have been performed on the Joint Fermilab - KICP Supercomputing Cluster,
supported by grants from Fermilab, Kavli Institute for Cosmological Physics, and the
University of Chicago. This work made extensive use of the NASA Astrophysics Data
System and {\tt arXiv.org} preprint server.

\bibliographystyle{apj}
\bibliography{citations,ak}

\end{document}